\documentclass[12pt,amsmath,amssymb,pre]{revtex4}
\usepackage{epsfig}
\newcommand{\map}{{\mathfrak M}}
\newcommand{\orb}{{\mathcal O}}
\newcommand{\invo}{{\mathfrak I}}
\newtheorem{theo}{{Theorem}}
\newtheorem{propo}{{Proposition}}
\newcommand{\Tr}{{\mbox{\rm Tr}}}
\newcommand{\rmap}{{\mathfrak R}}
\begin{document}
\title{Regimes of stability of accelerator modes}

\author{Rebecca Hihinashvili$^{1}$, Tali Oliker$^{1}$, Yaniv S. Avizrats$^{1}$, Alexander
Iomin$^{1}$, Shmuel Fishman$^{1}$, Italo Guarneri$^{2,3,4}$\\
{\small $^1$ Department of Physics, Technion, Haifa 32000, Israel}\\
{\small $^2$ Centro di Ricerca per i Sistemi Dinamici}\\
{\small Universit\`a dell'Insubria a Como, via Valleggio 11, 22100 Como, Italy}\\
{\small $^3$  Istituto Nazionale per la Fisica della Materia,
via Celoria 16, 20133 Milano, Italy}\\
{\small $^4$ Istituto Nazionale di Fisica Nucleare, Sezione di
Pavia, via Bassi 6, 27100 Pavia, Italy}} \vspace{.2cm}
\begin{abstract}
{The phase diagram of a simple area-preserving map, which was
motivated by the quantum dynamics of cold atoms , is explored
analytically and numerically. Periodic orbits of a given winding
ratio are found to exist within wedge-shaped regions in the phase
diagrams, which are analogous to the Arnol'd tongues which have
been extensively studied for a variety of dynamical systems,
mostly dissipative ones. A rich variety of bifurcations of various
types are observed, as well as period doubling cascades. Stability
of periodic orbits is analyzed in detail.}
\end{abstract}
\maketitle
\section{Introduction\label{introduction}}
\def\theequation{I. \arabic{equation}}

Periodic orbits are the backbone of the description of dynamical
systems \cite{book,Lichtenberg_Lieberman,Ott}, as they enable
global understanding of  motion in the entire phase space.  In the
present paper we explore the periodic orbits of the map \cite{GRF,
FGR1, FGR2}:
\begin{eqnarray}\label{I1}
J_{n+1}\;&=&\;J_n+\tilde{k}\sin\theta_{n+1}+2\pi\Omega~~~~\mbox{mod
$2\pi$}
\nonumber \\
\theta_{n+1}\;&=&\;\theta_n+J_n~~~~\mbox{mod $2\pi$}\;.
\end{eqnarray}
This map will be denoted $\mathfrak M$ in this paper. Like all the
other maps which are introduced in this paper, it maps the 2-torus
onto itself; however, "mod $2\pi$" will always be left understood
in all equations which follow. The map $\map$ reduces to the
Standard Map \cite{Lichtenberg_Lieberman} for $\Omega=0$. When
$\tilde{k}=0$, for each rational value $j/p$ of $\Omega$ the map
has periodic orbits of period $p$. For small $\tilde{k}>0$ such
orbits still exist in a range of  $\Omega$ around $j/p$. This
range of $\Omega$ decreases with $p$. The region  in the
$(\tilde{k},\Omega)$ phase diagram, where periodic orbits of
period $p$ are found, has the shape of a wedge, with the tip on
the $\Omega$ axis ($\tilde{k}=0$). This wedge grows wider as
$\tilde{k}$ increases, and intersects other wedges. Furthermore,
as $\tilde{k}$ increases more orbits are generated, others lose
stability, and an extremely complicated $(\tilde{k}, \Omega)$
phase diagram is generated (see Fig. 1 of \cite{GRF}). The
purpose of the present paper is to understand this diagram.\\
In the case of dissipative systems, this sort of phase diagrams
have been studied extensively \cite{Arnold, Arnold2, kapral}. The
standard sine-circle map \cite{Arnold}:
\begin{equation}
\label{sine} \theta_{n+1}=\theta_{n}-K\sin(\theta_{n})-2\pi\Omega,
\end{equation}
is a paradigm in the study of mode-locking for dissipative maps.
If $K = 0$ and $\Omega$ is a rational number $j/p$ any trajectory
of the sine-circle map returns to its initial value (modulo
$2\pi$) after $p$ iterations. For $0<K<1$, mode-locking is
observed; that is, over a range of $\Omega$ values around $j/p$
(the {\it mode-locking interval}) a periodic trajectory with
rational winding number $j/p$ {\em persists}. This periodic orbit
attracts all other orbits asymptotically in time, so that all of
them eventually acquire this winding number. The widths of the
mode-locking intervals are exponentially small in $p$, and
increase with $K$  up to $K=1$. The regions thus formed in
$(K,\Omega)$ parameter space, terminating at $K=0$, $\Omega=j/p$,
are known as {\em Arnol'd tongues\/} \cite{Arnold2,kapral}. For
$K>1$ the tongues intersect, stability of orbits is lost, and new
orbits are generated. In the same way as the sine-circle map
(\ref{sine}) is representative of dissipative systems, the map
(\ref{I1}) is representative of area preserving ones. The wedges
in the phase space of (\ref{I1}) where the periodic orbits of
period $p$ are found will again be referred to as Arnol'd tongues;
however, it will be shown that the phase space structure, and the
structure of the tongues in particular,  is significantly
different. This is expected, because orbits of area-preserving
maps
cannot attract.\\
The map (\ref{I1}) was obtained in studies of the quantum dynamics
of cold atoms, which are kicked by a standing light wave, and are
at the same time accelerated by gravity \cite{FGR1, FGR2,
experiment2, experiment1}. The gravitational acceleration is
proportional to $\Omega$, the strength of the kicking to $\tilde
k$, and $\epsilon$ is the separation from quantum resonance, which
occurs when  the ratio between the kicking period and a natural
period of the center-of-mass motion of the atoms is an integer. It
was shown that $\epsilon$ plays the role of Planck's constant, and
the limit $\epsilon \rightarrow 0$ (while $\tilde{k}=k \epsilon$
is fixed) is a pseudo-classical limit that is useful to describe
the dynamics for small $\epsilon$ \cite{FGR1, FGR2, ASFGreview}.
In this limit the map (\ref{I1}) was obtained. Its periodic
orbits, and the stable islands around them, account for the
Quantum Accelerator Modes \cite{FGR1, FGR2}, which were observed
experimentally \cite{experiment1, experiment2, exp_center}.

Compared to that of dissipative systems, the phase diagram of
(\ref{I1}) is much more complicated, mainly because overlaps of
different tongues occur at arbitrarily small values of $\tilde k$.
Therefore, its numerical exploration demands high precision in
determining periodic orbits. A property of (\ref{I1}) which proves
of decisive help in this task is that (\ref{I1}), like the
Standard Map, can be written as a product :
\begin{equation}\label{I1*}
\map\;=\;\invo_B\invo_A\;,
\end{equation}
where the maps $\invo_A$,$\invo_B$ are defined as:
\begin{eqnarray}\label{I2}
\invo_A\,:\;(J,\theta)&\rightarrow&\;(-J,\theta+J)\;\nonumber\\
\invo_B\,:\;(J,\theta)&\rightarrow&\;(-J+{\tilde
k}\sin(\theta)+2\pi\Omega\,,\,\theta).
\end{eqnarray}
and are involutions, that is $\invo_A^2=\invo_B^2=1$ (the identity
map). If a map has property (\ref{I1*}), then the search for its
periodic orbits is greatly simplified, if restricted to the class
of those orbits which include fixed points of either involution,
as we review in detail in Appendix \ref{AppInv}. This method was
applied in \cite{Greene} for the calculation of the last
separating KAM torus of the Standard Map. In the present work this
method is used to calculate periodic orbits of (\ref{I1}), both
analytically, in the perturbative regime in Sec \ref{pert}, and
numerically in other regimes in Sec.\ref{numerical}.

Many classic items of the theory of dynamical systems, such as
period-doubling cascades, and other bifurcations \cite{Ott, book,
Lichtenberg_Lieberman} are met in the study of map (\ref{I1}). In
our analysis we have focused on tongues where $\Omega$ at the tip
(vertex) takes the values $1$ (or $0$), $1/2$ and $1/3$. We
searched for periodic orbits such that one of their points is a
fixed point of one of the involutions.
As the boundaries of the tongue are crossed from inside, periodic
orbits are annihilated as a result of coalescence of a stable
orbit with an unstable one. Near the tip of the tongue $j/p$ the
orbits have $j$ and $p$ coprime. As $\tilde{k}$ is increased these
become unstable at some stability border where a period doubling
bifurcation takes place. The ratio $j/p$ does not change as a
result of continuity. This period doubling is followed by a period
doubling cascade, which satisfies a characteristic scaling rule of
such a route to chaos for area preserving systems, and is
universal \cite{mackay}. Also below the stability border a
pitchfork bifurcation of an orbit of period $p$ to two orbits of
the same period takes place, resulting in a cusp in the
$(\tilde{k},\Omega)$ phase diagram (compare to \cite{kapral}). The
analysis is performed analytically for small $\tilde{k}$ in
Sec.\ref{pert} and mostly numerically in Sec. \ref{numerical}.
Examples of periodic orbits that cannot be calculated with the
involution method are presented in Sec. \ref{noinvo}. The
involution method is reviewed in Appendix \ref{AppInv} and various
symmetries of the periodic orbits are presented in this Appendix
as well as in Appendix \ref{appsymmetry}. Properties of Gauss sums
and the relation to points on periodic orbits are outlined in
Appendix \ref{appgauss}. The numerical method for finding periodic
orbits is presented in Appendix \ref{appmatlab} and a Hopf
bifurcation is examined analytically in Appendix
\ref{apphopfbifu}. As tongues intersect, there are regions in the
$(\tilde{k},\Omega)$ phase diagram where orbits characterized by
different $j/p$ coexist. These orbits are found in different parts
of the $(\theta,J)$ phase
space, and typically do not "interact". \\
Here we are interested mainly in stable periodic orbits. If the
issue of stability is ignored, strong rigorous results can be
obtained. In particular, work by T. Oliker and B. Wajnryb
\cite{tali} implies that if a line in the tongue $(p_1,j_1)$
starts from its tip (vertex) and intersects the boundary of a
tongue $(p_2,j_2)$ which is its Farey neighbor [4,17-19], then it
intersects along the way all the tongues $(\tilde{p},\tilde{j})$
which are in the interval $[\frac{j_1}{p_1},\frac{j_2}{p_2})$,
intersecting all its Farey neighbors in the correct order. The
same work also suggests that in the center of the tongue $(j,p)$
all the orbits of period $p\cdot l$ ($l$ integer) exist, and their
winding number is $j/p$.


\section{Perturbative calculation of
periodic orbits}\label{pert}
\def\theequation{II. \arabic{equation}}
\setcounter{equation}{0} In this section the periodic orbits of
the map (~\ref{I1}) are calculated to the first order in
$\tilde{k}$. Throughout this section $J_n$ and $\theta_n$ are
understood  to be on the $2\pi$ torus, notably they are taken mod
$2\pi$; moreover, $\Omega$ is also taken mod $1$. First we note
that the $n-$th iterate of the map (~\ref{I1}) at zeroth order in
$\tilde{k}$ is:
\begin{eqnarray}\label{J_n_0}
J_n\,&=&\,J_0+2\pi\Omega n\nonumber\\
\theta_n\,&=&\,\theta_0+\sum_{n'=0}^{n-1}J_{n'}=\theta_0+J_0n+\pi\Omega
n(n-1)\,.
\end{eqnarray}
In order to determine $J_n$ at the 1st order in $\tilde k$, only
the $0-$th order in $\theta$ is required, and so, on substituting
the 2nd eq. of (\ref{J_n_0}) in (\ref{I1}), and iterating $n$
times, we find:
\begin{equation}\label{J_n}
J_n=J_0+\tilde{k}\sum_{n'=1}^{n}\sin[\theta_0+(J_0-\pi\Omega)n'+\pi\Omega
n'^{2}]+2\pi\Omega n.
\end{equation}
Substitution of this equation in the second equation of
(~\ref{I1}) yields
\begin{equation}\label{theta_n}
\theta_n=\theta_0+J_0n+\pi\Omega
n(n-1)+\tilde{k}\sum_{n'=1}^{n}(n-n')\sin[\theta_0+(J_0-\pi\Omega)n'+\pi\Omega
n'^2].
\end{equation}
For a periodic orbit of period $p$ the following equalities hold:
\begin{eqnarray}\label{theta}
\theta_p-\theta_0\,&=&\,2\pi s\,\nonumber\\
J_p-J_0\,&=&\,2\pi j,
\end{eqnarray}
where the integers $s$ and $j$ are the winding numbers of $\theta$
and $J$ around the torus. At the $0-$th order in $\tilde{k}$,
\begin{equation}\label{Omega_0}
\Omega=\frac{j}{p}
\end{equation}
must hold, and so periodic orbits can be found only for rational
values of $\Omega$. For {\em nonvanishing} $\tilde{k}$, periodic
orbits can be found in an interval around each rational value of
$\Omega$.
In what follows we solve for periodic orbits for small
$\tilde{k}$, assuming  that $(\Omega-\frac{j}{p})$ is also small
(this assumption will be verified {\it a posteriori}). Therefore
we replace $\Omega$ by $\frac{j}{p}$ in the terms that are
proportional to $\tilde{k}$ in (\ref{J_n}) and (\ref{theta_n}).

In order to find a periodic orbit of period $p$ , we seek for a
point $(\theta_0,J_0)$ which is at once a point in this orbit and
a fixed point point of one of the involutions (\ref{I2}) (Appendix
\ref{AppInv}). The fixed points of $\invo_A$ satisfy
\begin{equation}\label{Jodd}
J_0=0
\end{equation}
while the fixed points of $\invo_B$ satisfy
\begin{equation}\label{Jeven}
J_0=\frac{1}{2}\tilde{k}\sin\theta_0+\pi\Omega
\end{equation}
or
\begin{equation}\label{Jeven2}
J_0=\frac{1}{2}\tilde{k}\sin\theta_0+\pi\Omega+\pi.
\end{equation}
Since the periodic orbits of the map $\map$ of (\ref{I1}) are
identical to those of its inverse, we find in Appendix
\ref{appsymmetry} that if in a periodic orbit of $\map$ there is a
point with momentum $J_i$, then in a periodic orbit of this map,
not necessarily the same one, there is also a point with momentum
$-J_i$. If there is only {\em one} periodic orbit which is stable
(or unstable), then the corollary at the end of Appendix
\ref{appsymmetry} implies that for each point with momentum $J_i$
there is also one with momentum $-J_i$ in this orbit, and so if
$p$ is odd then $J_0=0$ is in this periodic orbit. Therefore $J_0$
of (\ref{Jodd}) is in a periodic orbit of  odd period $p$. On the
other hand, for even $p$, (\ref{Jeven}) or (\ref{Jeven2}) is in
the periodic orbit. This is discussed in Appendix \ref{AppInv}.

From(~\ref{J_n}) and (~\ref{theta}) we find that in the leading
order in $\tilde{k}$ (or $\epsilon$) a point on a periodic orbit
of period $p$ satisfies
\begin{equation}\label{j/p_Omega}
2\pi(\frac{j}{p}-\Omega)=\frac{1}{2i}\frac{\tilde{k}}{p}\left\{
~\left(e^{i\theta_0}\sum_{n'=1}^{p}e^{i\pi
\frac{j}{p}[(\chi_p-1)n'+n'^2]}\right)-c.c.~ \right\}
\end{equation}
where $\chi_p=0$ for odd $p$, where (\ref{Jodd}) holds, while
$\chi_p=1$ for even p where (\ref{Jeven}) holds and $\chi_p=1+p/j$
where (\ref{Jeven2}) holds. We will solve equation
(\ref{j/p_Omega}) for $\theta_0$ to the first order in $\tilde{k}$
and then show that it is consistent with (\ref{theta_n}) in this
order.

For this purpose it is instructive to use properties of Gauss sums
\cite{GRF, Gauss sums, talbot, Gauss sums2, Dirichlet} . We
consider
\begin{equation}\label{Gpj}
G(p,j)=\sum_{m=1}^{p}e^{i\pi \frac{j}{p}[(\chi_p-1)m+m^2]}
\end{equation}
which can be summed (\cite{Gauss sums, talbot, Gauss sums2,
Dirichlet}, see Eq. (36) of \cite{GRF}) to give
\begin{equation}\label{Gpjcompact}
G(p,j)=\sqrt{p}e^{i\xi(j,p)}
\end{equation}
where $\xi(j,p)$ is real (see Appendix \ref{appgauss}).
Substitution in (\ref{j/p_Omega}) leads to
\begin{equation}\label{theta3}
2\pi\left(\frac{j}{p}-\Omega\right)=\frac{\tilde{k}}{\sqrt{p}}\sin\vartheta
\end{equation}
where
\begin{equation}\label{theta3def}
\vartheta=\theta_0+\xi(j,p).
\end{equation}
A $\theta_0$ that belongs to a periodic orbit must satisfy this
equality.

The Gauss sum (\ref{Gpj}) also satisfies (this is a corollary of
Eqs. (34) and (35) of {\cite{GRF})
\begin{eqnarray}\label{m*Gpj}
\sum_{m=1}^{p}m e^{i\pi \frac{j}{p}[(\chi_p-1)m+m^2]}=\left\{
\begin{array}{ll}
\frac{p+1}{2}G(p,j) & \mbox{p odd} \\

\\
\frac{p}{2}(G(p,j)+1) & \mbox{p even} \\
\end{array}
\right.~.
\end{eqnarray}
Substitution in (\ref{theta_n}) for $n=p$, where $p$ is odd, and
using (\ref{theta}) results in:
\begin{equation}\label{s_podd}
2\pi\left[\frac{s}{p}-\frac{\Omega}{2}(p-1)\right]=\frac{\tilde{k}}{\sqrt{p}}\frac{p-1}{2}\sin\vartheta.
\end{equation}
It is easy to see that it is consistent with (\ref{theta3}) for
$s=\frac{j(p-1)}{2}$. For even $p$ one finds in a similar way that
\begin{equation}\label{s_peven}
2\pi[\frac{s}{p}-\frac{\Omega}{2}p]=\frac{\tilde{k}\sqrt{p}}{2}\sin\vartheta
\end{equation}
and consistency with (\ref{theta3}) is possible for
$s=\frac{pj}{2}$. Equation (\ref{theta3}) determines the periodic
orbits in small $\tilde{k}$. One finds solutions in the range
\begin{equation}\label{range}
|\frac{j}{p}-\Omega|\leq\frac{\tilde{k}}{2\pi\sqrt{p}}~.
\end{equation}
This region has the form of a wedge and for small $\tilde{k}$ its
boundaries are the lines
\begin{equation}\label{edges}
\Omega=\frac{j}{p}\pm\frac{\tilde{k}}{2\pi\sqrt{p}}~,
\end{equation}
and they are presented in Fig. \ref{edgesfig}.


\begin{figure}
\centerline{\epsfxsize 14cm \epsffile{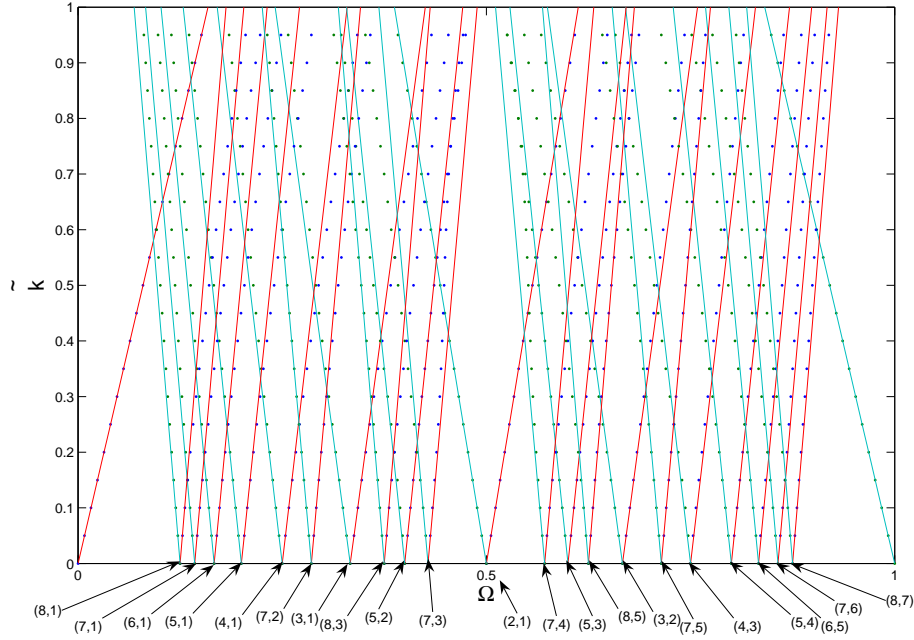}}
\caption{Various tongues, marked by $(p,j)$, in the $\tilde{k}$,
$\Omega$ phase diagram. For small $\tilde{k}$ the numerical
results (dots) coincide with the analytically determined tongue
boundaries (solid lines).} \label{edgesfig}
\end{figure}


Each tongue is characterized by $p$ and $j$. The inequality
(\ref{range}) justifies our assumption that $|\frac{j}{p}-\Omega|$
is at most of order $\tilde{k}$ (or $\epsilon$). We see that
indeed for small $\tilde{k}$ the boundaries are well described by
(\ref{edges}). The point $\vartheta$ on the orbit is given by
(\ref{theta3}). Special points are the two in the center of the
tongue $\vartheta=0,\pi$ and the two on its boundaries
$\vartheta=\frac{\pi}{2},\frac{3}{2}\pi$. Since $\xi(j,p)$ is
known, one can determine $\theta_0$ from $\vartheta$, as will be
shown in Appendix \ref{appgauss} (see table \ref{ksitable}). The
specific values of $\theta_0$ are of no importance for discussion
in the rest of this section, since all the results may be
expressed in terms of $\vartheta$. In the regime of small $\tilde
k$, for each value of $\tilde{k}$ and $\Omega$ there are two
periodic orbits, one is stable and the other is unstable. The two
get closer and closer as the boundary is approached and coalesce
at the tongue boundaries. We tested numerically the validity of
(\ref{theta3}) for small $\tilde{k}$.\\
The stability of an orbit can be determined as usual from the
tangent map. The tangent map of (\ref{I1}) is
\begin{equation}\label{T}
T(\theta)= \left ( \begin{array}{cc}
1+\tilde{k}\cos\theta & \tilde{k}\cos\theta \\
1 & 1
\end{array} \right )
\end{equation}
and the tangent map over $p$ iterates along a period$-p$ orbit
(also termed Monodromy matrix) is given by
\begin{equation}\label{Ttheta1}
T_p(\theta_0)=\prod_{i=0}^{p-1}T(\theta_i)~.
\end{equation}
A periodic orbit is stable if
\begin{equation}\label{Tr2}
|Tr[T_p(\theta_0)]|< 2
\end{equation}
and unstable if the opposite inequality holds. In order to
calculate the product (\ref{Ttheta1}) to the first order in
$\tilde{k}$ we note that (\ref{T}) can be decomposed as
\begin{equation}\label{decomT}
T(\theta_{i})=\overline{T}_0+\tilde{k}\overline{T}_1(\theta_{i})
\end{equation}
where
\begin{equation}\label{T0}
\overline{T}_0=\left( \begin{array}{cc}
1 & 0 \\
1 &1
\end{array}
\right)
\end{equation}
and
\begin{equation}\label{T1}
\overline{T}_1(\theta_i)=\left( \begin{array}{cc}
\cos\theta_i & \cos\theta_i \\
0 & 0
\end{array}
\right).
\end{equation}
With the help of the identities $\Tr(A+B)=\Tr A+\Tr B$ and
$\Tr(AB)=\Tr(BA)$ we find that, up to first order in $\tilde{k}$
\begin{equation}\label{TrTp}
\Tr(T_p(\theta_0))=\Tr\overline{T}_0^p+\tilde{k}\sum_{i=1}^p\Tr(\overline{T}_0^{(p-1)}\overline{T}_1(\theta_i)).
\end{equation}
It is easy to see that
\begin{equation}\label{T0p}
\overline{T}_0^p=\left( \begin{array}{cc}
1 & 0 \\
p & 1
\end{array}
\right)
\end{equation}
and
\begin{equation}\label{T0T1}
\overline{T}_1(\theta_i)\overline{T}_0^{(p-1)}=\left(
\begin{array}{cc}
p\cos\theta_i & \cos\theta_i \\
0 & 0
\end{array}
\right).
\end{equation}
Using (\ref{Gpj}),(\ref{Gpjcompact}), and  the fact that the
angles along a periodic orbit are given to the zeroth order in
$\tilde{k}$ by the 2nd equation in (\ref{J_n_0}), one finds that
\begin{equation}\label{Trace}
\Tr({T}_p(\theta_0))=2+\tilde{k}p^{3/2}\cos\vartheta
\end{equation}
where $\vartheta$ is related to $\theta_0$ by (\ref{theta3def}).

At the boundaries of the tongue given by (\ref{edges}) the stable
and unstable orbits coalesce and $\Tr({T}_p(\theta_0))=2$. Inside
the tongue the orbits with
$\frac{\pi}{2}<\vartheta<\frac{3}{2}\pi$ are stable and the ones
in the other part are unstable. This was numerically tested.

\section{Numerical analysis of periodic orbits and stability borders}
\label{numerical}}
\def\theequation{III. \arabic{equation}}
\setcounter{equation}{0}

In the previous section the $\tilde{k},\Omega$ phase diagram was
analyzed in the framework of perturbation theory for small
$\tilde{k}$. This approach is valid near the tip of a tongue. In
this section the phase diagram is studied in the non-perturbative
region. In this region, the edges of a tongue deviate from the
straight lines (\ref{edges}) (except in the case $p=1$, where
(\ref{edges}) is exact). Furthermore,  this region is
characterized by instabilities and bifurcations, which are
observed  as $\tilde{k}$ is increased. We numerically explore this
behavior for
some low-period tongues.\\
As periodic orbits coalesce at the edges of a tongue, their
precise numerical determination is difficult near the edges; and
this, in turn, makes the precise determination of the edges
themselves somewhat delicate. An algorithm we have developed to
locate the boundary is presented in Appendix \ref{appmatlab}. From
(~\ref{theta3}) one finds that if $\vartheta_s$ and $\vartheta_u$
belong to the stable and unstable orbits respectively, then, in
the vicinity of the boundary, and for small $\tilde k$,
\begin{equation}\label{u_s}
(\vartheta_u-\vartheta_s)^2=16\pi\sqrt{p}\frac{|\delta\Omega|}{\tilde{k}}
\end{equation}
where $\delta\Omega$ is the distance from the boundary in
$\Omega$. This equation can be used to check consistency of
results.

The region of existence of a {\it stable} $(j/p)$ orbit is bounded
on the left and on the right  by the edges of the tongue, and from
above by a stability border, which lies in the interior of the
tongue ( see, e.g. Fig. \ref{p1tong} ). Whereas at the edges of
the $(j/p)$ tongue, period $p$ orbits, whether stable or unstable,
disappear, at the stability border stable periodic orbits become
unstable, and, in the
cases considered in this work, a period-doubling cascade follows.\\
Loss of  stability is foreshadowed by the perturbative equation
(\ref{Trace}), because the second term in it is negative for the
stable orbit, and so it would lead to violation of (\ref{Tr2}),
and hence to instability, for
$|\tilde{k}p^{3/2}\cos(\vartheta)|>4$. According to this argument,
the location of the stability boundary in $\tilde{k}$ decreases
with the period as $p^{-3/2}$ (in agreement with earlier work
\cite{GRF}). However, this instability occurs for relatively large
values of $\tilde{k}$, where perturbation theory is no longer
valid, and therefore (\ref{Trace}) may just provide hints there .
In particular, it suggests that stable regions are more likely to
be found close to the margins of a tongue, where $\cos(\vartheta)$
is small. This is indeed observed in numerical results: as $\tilde
k$ increases, the stability border asymptotically approaches the
edges of the tongue.
\\
All periodic orbits which are calculated in this section are found
by the involution method, which is described in Appendix
\ref{AppInv}. As explained in Appendix \ref{appmatlab}, they are
numerically found from the requirement that one point in the orbit
is a fixed point of an involution. In the case of odd period $p$,
this is involution $\invo_A$ of (\ref{I2}); for even $p$, it  is
typically  $\invo_B$ of (\ref{I2}).

\subsection{The tongue $j=1,\; p=1$. \label{IIIA}}

\begin{figure}
\begin{center}
\epsfxsize=15.6cm \leavevmode
    \epsffile{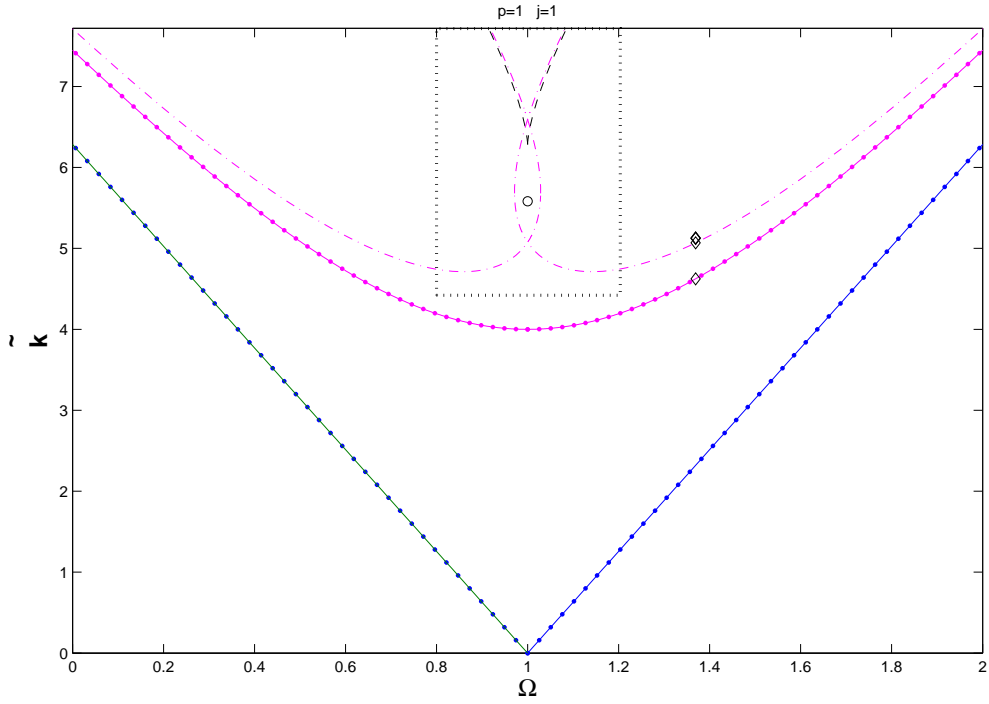}
\caption{Phase diagram of the $(j=1, p=1)$ tongue. Full lines show
the analytical edges of the tongue (straight lines emanating from
the point marked by $1$ on the horizontal axis) and the analytical
stability border of period 1 orbits. Superimposed dots show
numerical results. The meaning of symbols (circle, rhombi) is
explained in the text. The dash-dotted line marks the stability
border of the $p=2$ orbit which emerges at the period doubling of
the $p=1$ orbit. The region delimited by the dotted-line rectangle
is described in Fig. \ref{p1zoom}. }\label{p1tong}
\end{center}
\end{figure}

\begin{figure}
\begin{center}
\epsfxsize=15.6cm \leavevmode
    \epsffile{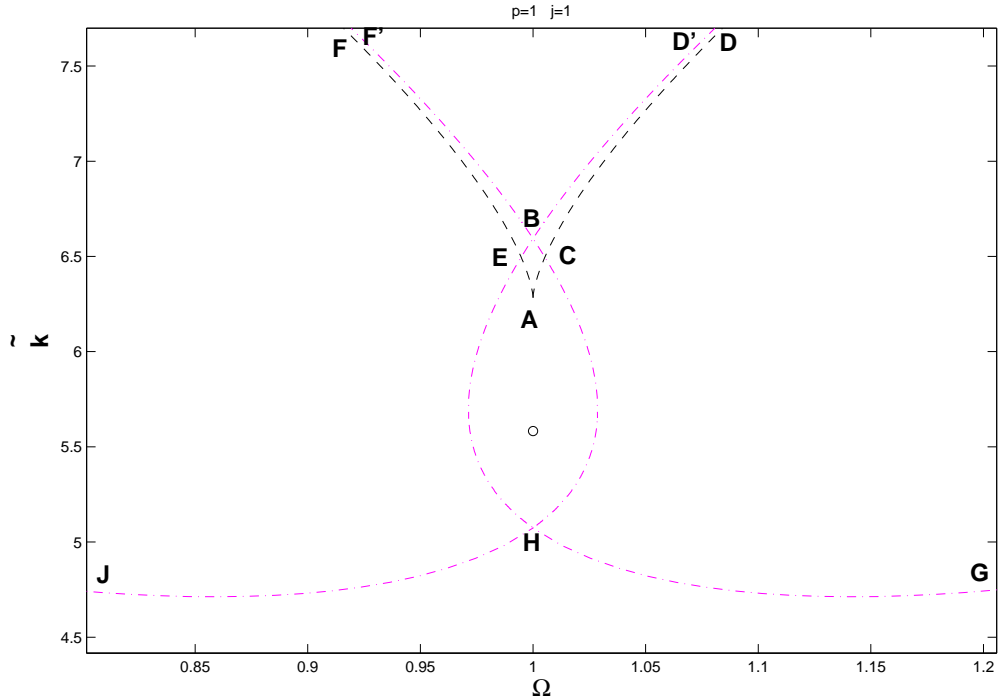}
\caption{A zoom-in of the region delimited by the rectangle in
Fig. \ref{p1tong}. Stable period 2 orbits are found below the line
JHG and inside the loop BHB, as well as between the lines EF and
BF', and between BD' and CD. Inside the kite-shaped region AEBC,
two stable orbits of period 2 are found. A very complicated
behaviour is found in the vicinity of H where period 2 and period
4 stable orbits coexist.}\label{p1zoom}
\end{center}
\end{figure}

For this tongue many results are known analytically, and were
presented in \cite{FGR1, FGR2, ASFGreview}. In particular the
boundary of the tongue, and the stability border satisfy
(\cite{FGR2}, Eq. (32)):
\begin{eqnarray}
\tilde{k}_b\;&=&\;2\pi|\Omega -1|\label{Kb}\;,\\
\tilde{k}_s\;&=&\;\sqrt{16+4\pi^2(\Omega-1)^2}\;.\label{Ks}
\end{eqnarray}
These are shown in Fig. \ref{p1tong} along with numerically
obtained data (the latter are shown in order to check consistency
of our numerical method of finding the boundary). As  ${\tilde k}$
increases beyond $\tilde{k_s}$ a period doubling bifurcation takes
place, such that the period 1 orbit turns unstable and a period 2
stable periodic orbit appears.  Around the $p=2$ orbit a resonant
orbit of period $2\cdot 3$ is formed, as  is shown in Fig.
\ref{port1} for the case marked by a circle in Fig.\ref{p1tong}.
This mechanism is discussed in Appendix \ref{apphopfbifu}. For
$\Omega=1$, $\tilde{k}=2\pi$ a pitchfork bifurcation of the $p=2$
orbit into two $p=2$ orbits takes place. The regions where such
orbits exist are bound by dashed lines in Fig.\ref{p1tong}. They
are magnified in Fig. \ref{p1zoom} and described in detail in
the corresponding caption. \\
The bifurcation which is observed on crossing the stability border
upwards is actually the 1st in a period-doubling cascade (marked
in Fig. \ref{p1tong} at $\Omega=1.3692$ by rhombi). Let us denote
by $\delta\theta_n$ the difference in $\theta$ between the points
separated by half a period in the orbit of period $2^np$, when it
becomes unstable; $p$ is the original period at the start of the
period-doubling cascade.
 We follow the cascade up to $n=4$ and find that
the numerical results as $n$ increases are consistent with:
\begin{equation}\label{delta}
\tilde{k}_n-\tilde{k}_{\infty}\sim\frac{1}{\delta ^n}
\end{equation}
\begin{equation}\label{alpha}
\delta\theta_n\sim\frac{1}{\alpha ^n}
\end{equation}
where ${\tilde k}_n$ are values of $\tilde k$ where the $(n+1)-$th
bifurcation is observed. Estimates for $\delta$ and $\alpha$ are
reported in Table \ref{tav2}, which summarizes this sort of
estimates for other tongues as well.
$\tilde{k}_{\infty}$ was calculated using the ratio between the
LHS of (\ref{delta}) for $n$ with the one for $n+1$, thus
eliminating the proportionality coefficient. The scaling factors
$\delta$ and $\alpha$ are universal numbers for the
period-doubling route to chaos, in area preserving systems with
2-dimensional phase space. The theoretical prediction of the
renormalization group treatment is $\delta\approx 8.721$
$\alpha\approx -4.018$ \cite{mackay}.

\begin{figure}
\begin{center}
\epsfxsize=12.6cm \leavevmode
    \epsffile{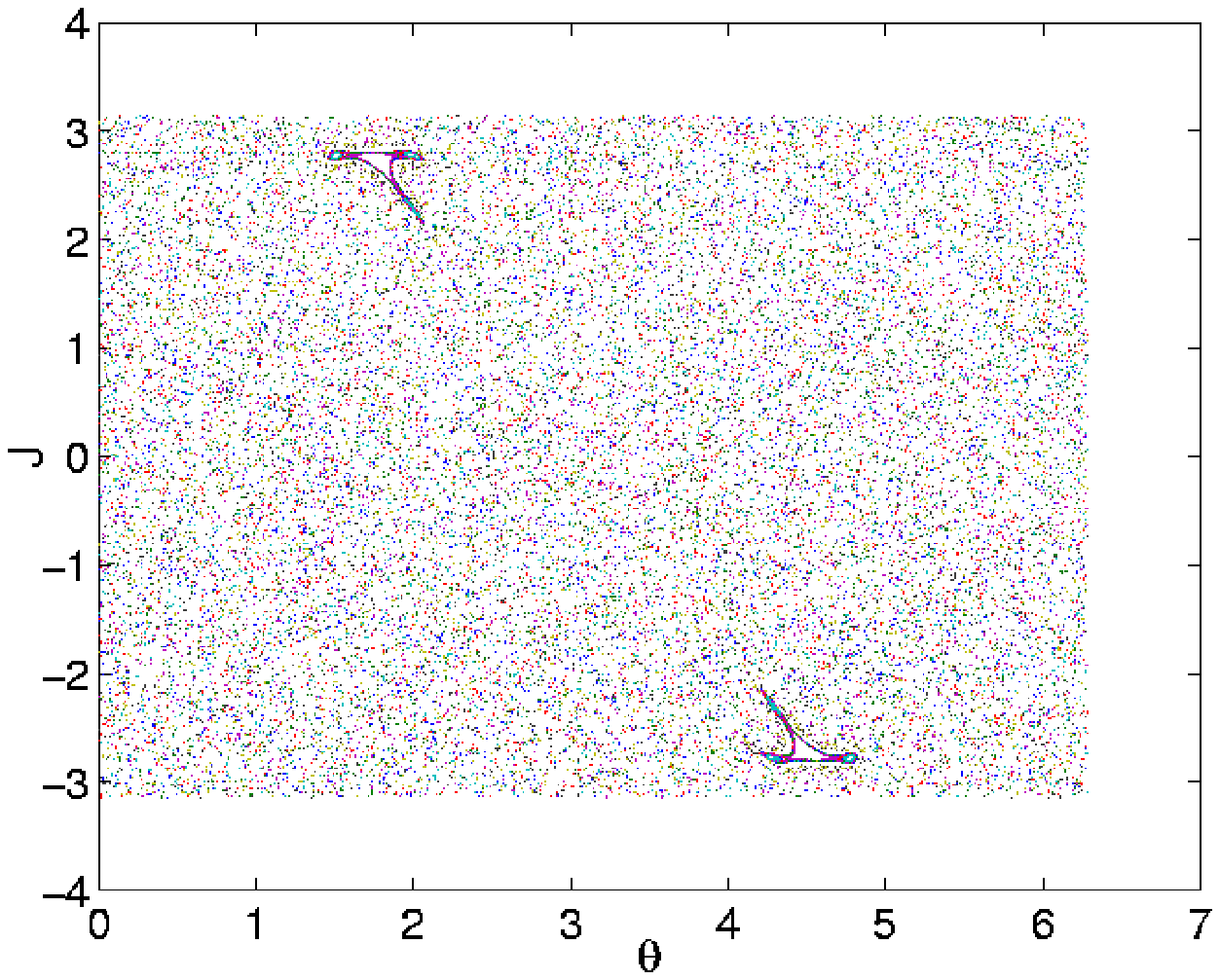}
\caption{Phase portrait for $\tilde{k}=5.5825>\tilde{k}_s$, and
$\Omega=1$.}\label{port1}
\end{center}
\end{figure}

\subsection{The tongue $j=1\;,p=2$.}

\begin{figure}
\begin{center}
\epsfxsize=15.6cm \leavevmode
    \epsffile{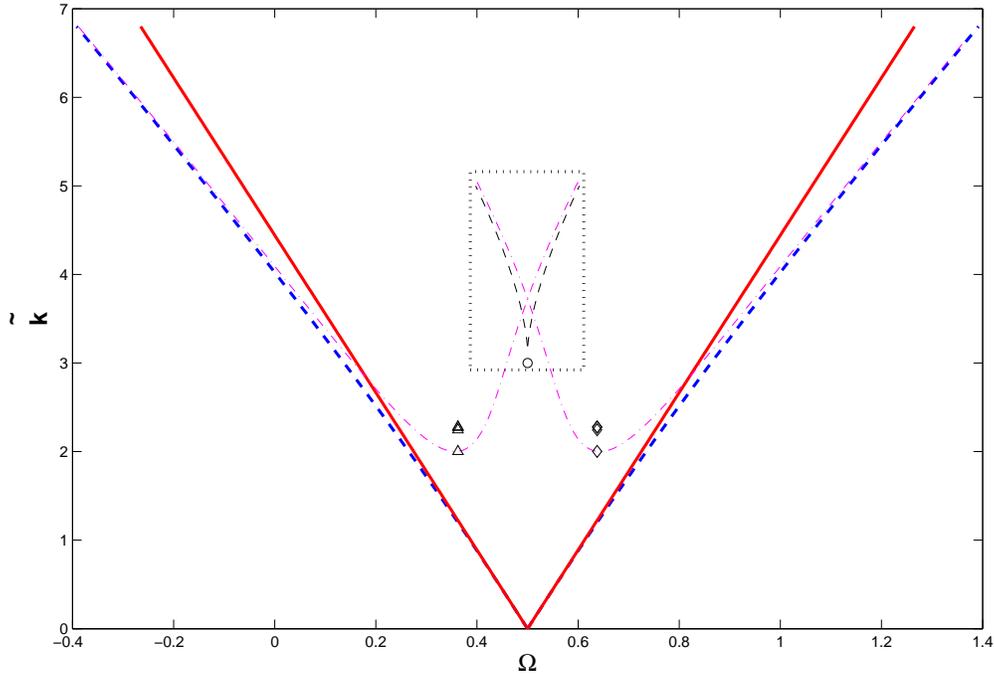}
\caption{Phase diagram for the $p=2, ~j=1$ tongue. The region of
coexistence of two stable periodic orbits of period $2$ is framed
by the dotted-line rectangle, and zoomed in Fig. \ref{p2zoom}. The
thick dashed lines mark the tongue boundaries (found numerically),
and the dash-dotted line is the stability border (found
numerically). The solid lines indicate the perturbative analytical
result of Eq. (\ref{edges}) for the boundary of the tongue. The
small rhombi and triangles indicate the period doubling cascade
for $\Omega=0.6376$ and $\Omega=0.3624$ correspondingly.
}\label{p2tong}
\end{center}
\end{figure}

\begin{figure}
\begin{center}
\epsfxsize=15.6cm \leavevmode
    \epsffile{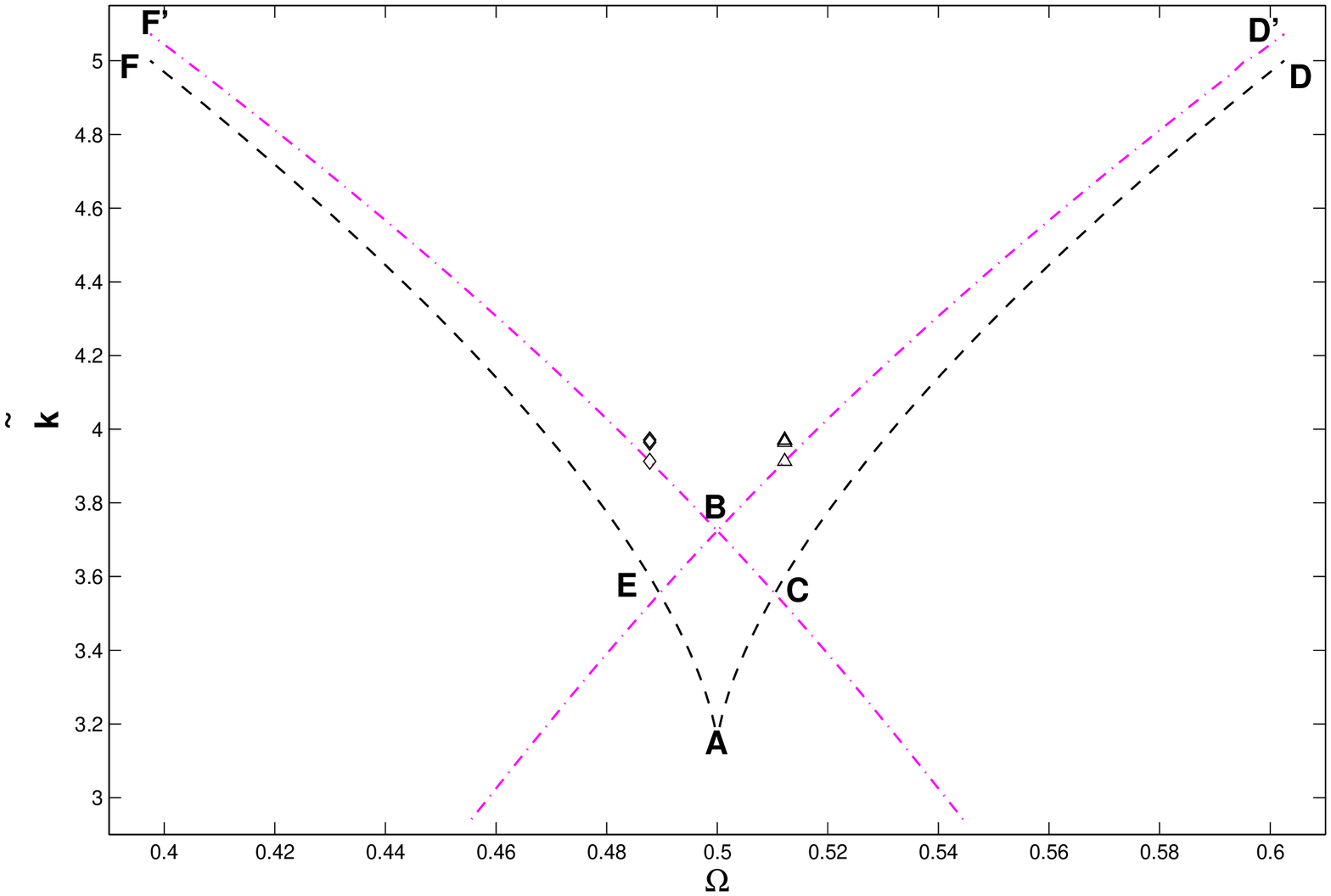}
\caption{A zoom-in of the region delimited by the dashed-line
rectangle of Fig.\ref{p2tong}. At the dash-dotted lines period-2
orbits lose stability, and stable period-4 orbits are formed. On
the dashed lines coalescence of one of the period-2 orbits  with
an unstable periodic orbit occurs. This stability loss at the
dash-dotted lines is followed by a period doubling cascade. The
small rhombi and triangles indicate this cascade for
$\Omega=0.5122$ and $\Omega=0.4878$ correspondingly.
}\label{p2zoom}
\end{center}
\end{figure}

\begin{figure}
\begin{center}
\epsfxsize=12.6cm \leavevmode
    \epsffile{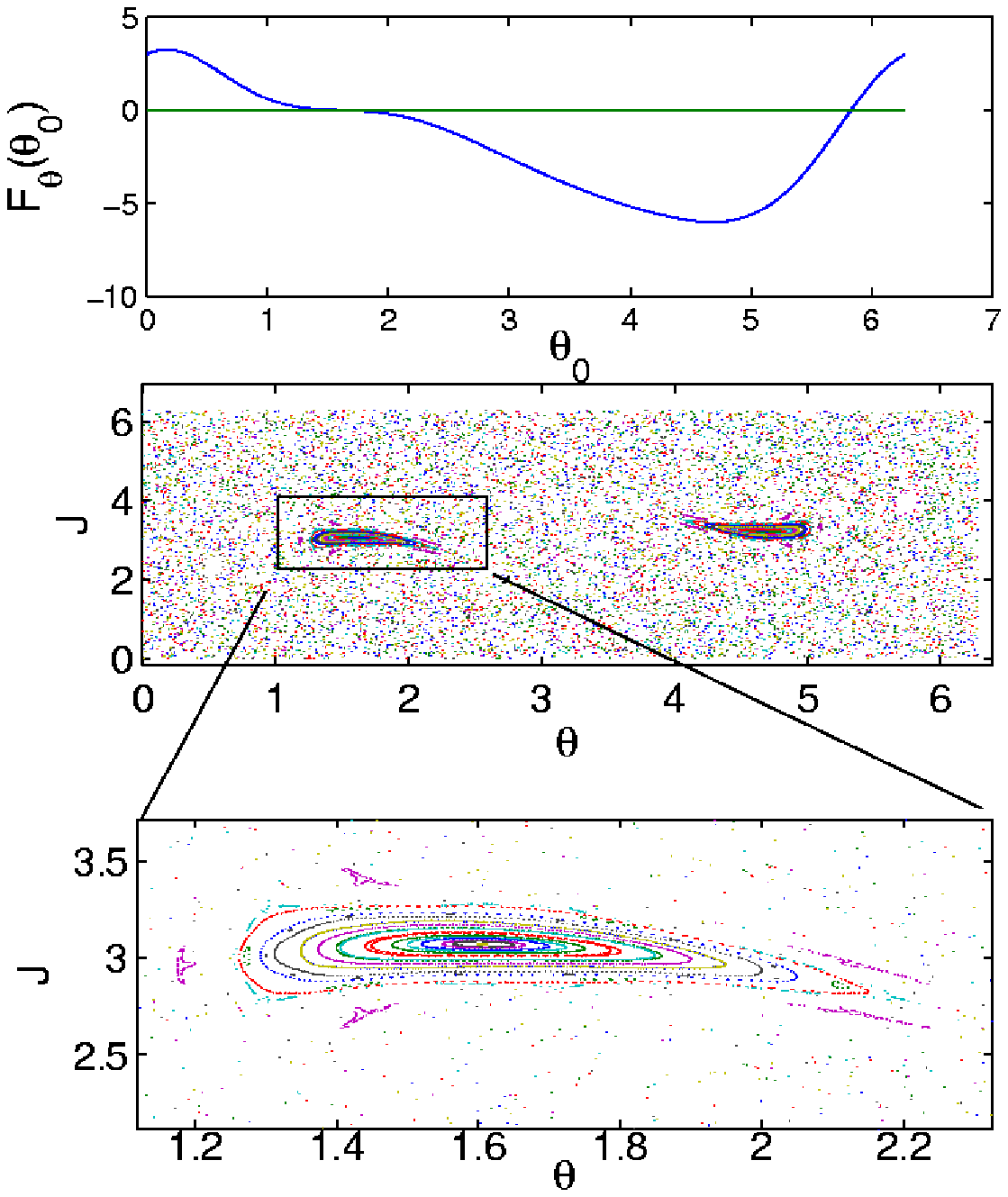}
\caption{Phase portrait for $p=2,~j=1$, $\tilde{k}=3.0$ and
$\Omega=1/2$. The upper slice presents the function
$F_{\theta}(\theta_0)$ . The middle slice shows the period two
stable orbit and the lower slice is a zoom on one of its
island.}\label{port21}
\end{center}
\end{figure}

The structure of this tongue is presented in Figs. \ref{p2tong}
and \ref{p2zoom}. First we study the phase portrait at the center
of the tongue ($\Omega=1/2$) for various values of $\tilde{k}$.
For $\tilde{k}=3.0$ (marked by "o" in Fig. \ref{p2tong}), which is
below the stability line , the phase portrait  is presented in
Fig. \ref{port21} (2nd and 3d slice from top). In the upper slice,
the function $F_{\theta}(\theta_0)$ is plotted, the zeros of which
are used to locate $\theta_0$ of periodic orbits according to the
Involution method, as explained in Appendix \ref{appmatlab}.
A similar  analysis was performed for all periodic orbits. As we
increase $\tilde{k}$, but keep it lower than the stability border
$\tilde{k}_s$, a new stable orbit of period two appears at
$\tilde{k}\geq\pi$. Further  bifurcations are found for
$\tilde{k}>\pi$, and these are described in Fig. \ref{p2zoom}.


The bifurcation leading from one stable period-2 orbit to two
distinct stable orbits with period 2, which is numerically
observed at $\tilde{k}=\pi$, is easily obtained analytically. It
is a standard pitchfork bifurcation. To see this, we iterate the
map (\ref{I1}) twice. The condition for a period two orbit (modulo
$2\pi$) is
\begin{equation}\label{pitch1}
\tilde{k}\sin\theta_0+\tilde{k}
\sin\left(\theta_0+\frac{\tilde{k}}{2}\sin\theta_0+\frac{\pi}{2}\right)=0,
\end{equation}
where $\Omega=1/2$ and (\ref{Jeven}) were used.
When $\tilde{k}\leq\pi$ there is one stable trajectory. At
$\tilde{k}=\pi$, this trajectory is $\{(\pi/2,\pi)\;,\;
(3\pi/2,\pi)\}$ and is marginally stable  (the trace of the
linearized map ${\rm Tr} [T_2(\theta_0)]=2$). When
$\tilde{k}=\pi+\delta\tilde{k}$, two new stable trajectories
emerge. To find these zeros of Eq. (\ref{pitch1}) we replace
$\theta_0=\pi/2+\delta\theta_0$ and neglect terms of order of
$O(\delta\tilde{k}{\delta\theta_0}^2)\sim O({\delta\theta_0}^4)$
and higher. This yields the following cubic equation for
$\delta\theta_0$:
\begin{equation}\label{patch2}
\delta\theta_0\cdot\left(\sqrt{\frac{2\delta\tilde{k}}{\pi}}+\delta\theta_0\right)\cdot
\left(\sqrt{\frac{2\delta\tilde{k}}{\pi}}-\delta\theta_0\right)=0\,
.
\end{equation}
The same equation is obtained from the substitution $\theta_0
=3\pi/2+\delta\theta_0$. Eq. (\ref{patch2}) yields three periodic
orbits with period 2. One of them has $\delta\theta_0=0$. It
continues the unique orbit at ${\tilde k}=\pi$ and, like that
orbit, it is marginally stable {\it at this order}; numerically,
it was found that it is actually unstable, and that
$0\neq\delta\theta_0=o(\delta\tilde{k})$. The other two orbits
have $\delta\theta_0=\pm\sqrt{\frac{2\delta\tilde{k}}{\pi}}$. They
are stable at this order, sharing the same value of ${\rm Tr}[
T_2]=2-2\pi\delta\tilde{k}$, and are numerically found to be
really stable.
\\
On further moving upwards along the axis of the tongue ( that is,
increasing $\tilde k$ while $\Omega=1/2$), the two period-2 stable
orbits persist until point $B$ is reached. There they turn
unstable, and give rise to two stable period-4 orbits. Two stable
period-2 orbits (which we term "$a$" and "$b$" for convenience)
are actually observed inside the whole kite-shaped region $AEBC$
in Fig.\ref{p2zoom}. Each of them is observed , alone, in a larger
region. For instance , "$a$" is observed in between the lines
$AD$, $ED'$, and "$b$" is is observed inbetween $AF$ and $CF'$ .
On the line $AD$ , "$a$" coalesces with an unstable orbit and
disappears,  and "$b$" disappears in the same way along the line
$AF$. Across the line $ED'$, "$a$" bifurcates to a period-4 orbit,
and the same happens to "$b$" across $CF'$.
Resonances of order 3 similar to those found for the $p=1$ tongue
and are presented in Fig. \ref{port1} are found also here . As
$\tilde{k}$ is increased above the stability border (dot-dashed
line in Fig. \ref{p2tong}), a period doubling cascade is started.
This was studied in the same way as for the $p=1$ tongue. Results
are shown in Table \ref{tav2}, and are marked by rhombi and
triangles in Figs. \ref{p2tong} and \ref{p2zoom}.

So far the periodic orbits in the tongue $p=2$, $j=1$ were found
identifying one point of the orbit as a fixed point of $\invo_B$.
There is also at least one stable periodic orbit that none of its
points are fixed points of $\invo_B$, but some are fixed points of
$\invo_A$. For $\Omega =1/2$ one such an orbit of period 4 is
$\{X_0=(0,0)\;,\,X_1=(\pi,0)\,,\, X_2=(0,\pi)\,,\,X_3=(\pi
,\pi)\}$, as is easily  checked . The points $X_0$ and $X_2$ are
fixed points of $\invo_A$ while $X_1$ and $X_3$ are interchanged
by this involution. All this demonstrates the Corollary in
Appendix \ref{AppInv}. We checked (by calculating the trace of the
tangent map) that the orbit is stable for $0<\tilde{k}\leq 0.7320$
and $2.7320\leq\tilde{k}\leq 2.8284$.


\subsection{The tongue $p=3,~j=1$}

\begin{figure}
\begin{center}
\epsfxsize=15.6cm \leavevmode
    \epsffile{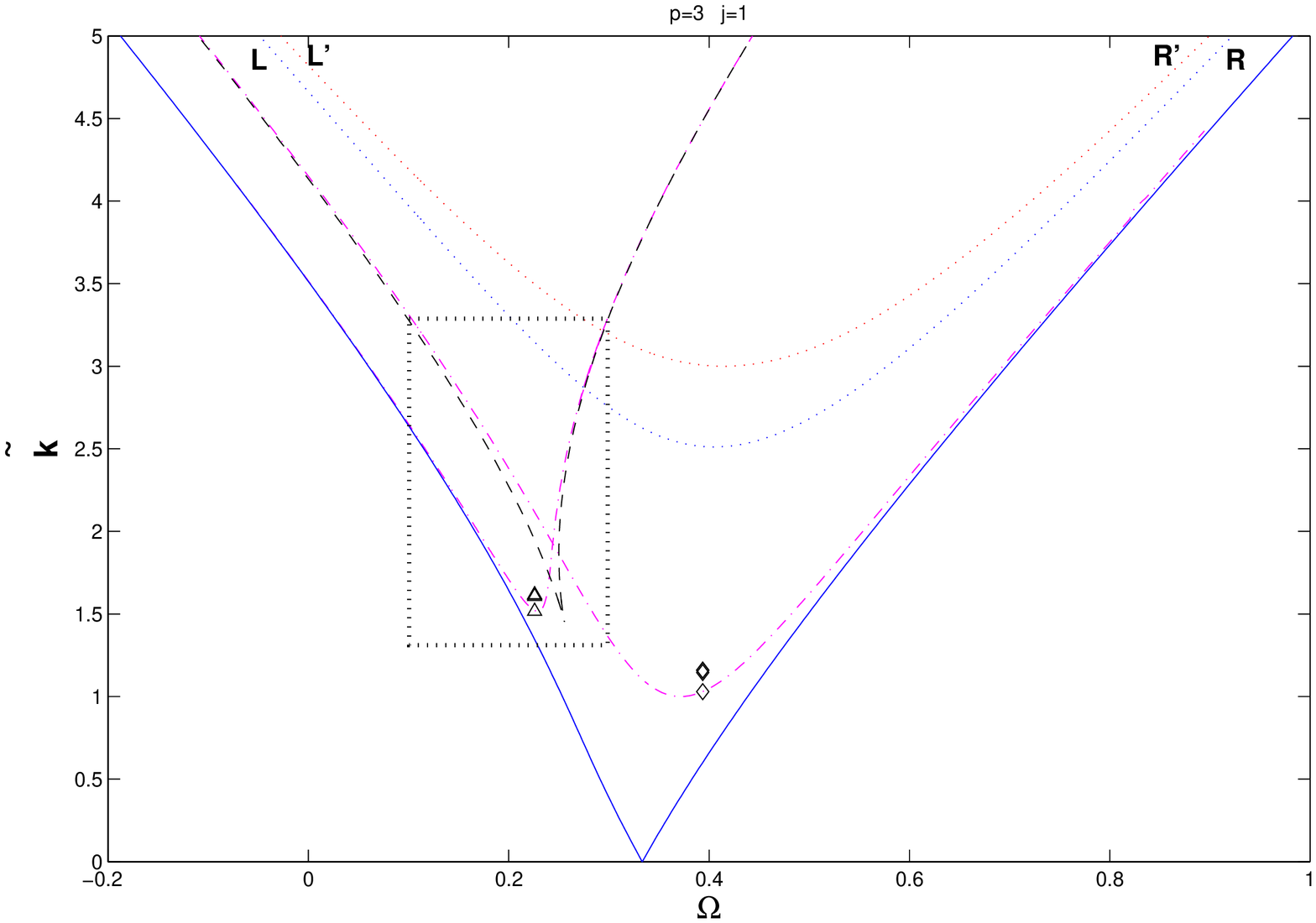}
\caption{Phase diagram for the $p=3, ~j=1$ tongue, indicating the
region of coexistence of two stable period-3 orbits,  which is
zoomed in Fig. \ref{p3zoom}. The solid line marks the tongue
boundaries found numerically, and the dash-dotted line is the
stability border also found numerically. The dashed lines bound
the region in which two period-3 orbits coexist. As each of these
lines is crossed from the inside of this region, one periodic
orbit coalesces with an unstable orbit, and is annihilated. At the
stability border a period doubling cascade begins. The rhombi and
triangles illustrate such a cascade for $\Omega=0.3936$ and
$\Omega=0.2285$ correspondingly.
 Between lines $LR$ and $L'R'$ a stable period 3 orbit is again observed.}
 \label{p3tong}
\end{center}
\end{figure}
\begin{figure}
\begin{center}
\epsfxsize=15.6cm \leavevmode
    \epsffile{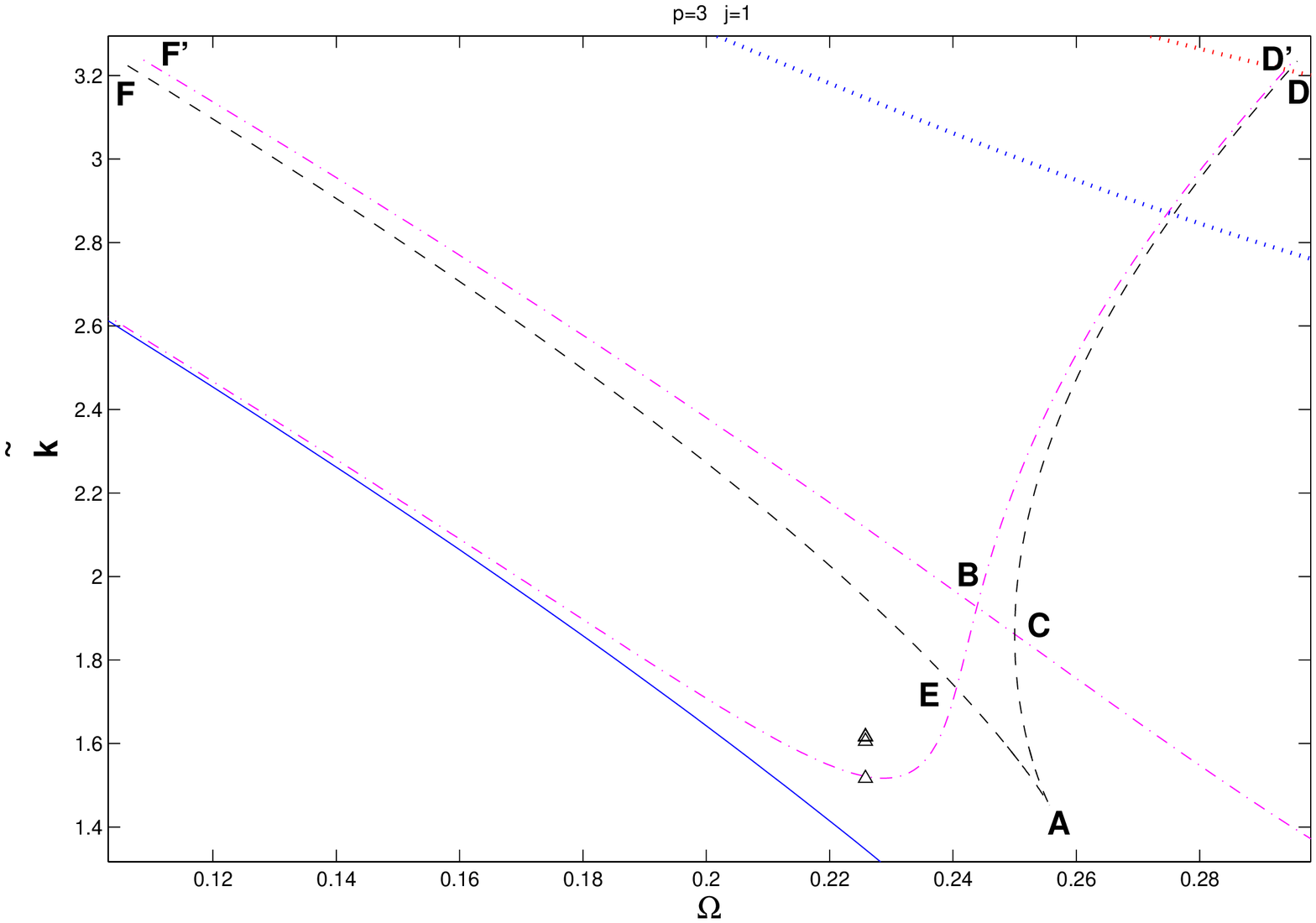}
\caption{A zoom-in of the region delimited by the rectangle in
Fig. \ref{p3tong}.}\label{p3zoom}
\end{center}
\end{figure}

Our numerical findings about the $p=3$, $j=1$ tongue are
summarized by Figs. \ref{p3tong} and \ref{p3zoom}.

The stable period-3 orbit is lost going upwards across the
stability border, the details of which are shown in Figs.
\ref{p3tong} and \ref{p3zoom} . However, a stable such orbit
reappears on further increasing $\tilde k$; it is observed in the
region bound by the lines $LR$ and $L'R'$. Between $A$ and $B$ in
Fig. \ref{p3zoom} there exist two stable periodic orbits of period
$3$ which we call "$a$" and "$b$". Both turn unstable at $B$.
"$a$" coalesces with an unstable orbit and disappears on the line
$AD$, and "$b$" disappears in the same way along $AF$. When
crossing $CF'$, "$b$" loses stability, and the same happens to
"$a$" along $ED'$.


\section{Orbits that are not related to involutions.}\label{noinvo}
\def\theequation{IV. \arabic{equation}}

In the previous sections periodic orbits that include fixed points
of $\invo_A$ or $\invo_B$ were found. An obvious question is
whether periodic orbits can be found, that do not include such
fixed points at all. First we find such orbits for $\tilde{k}=0$.
For $\Omega=j/p$, a point $(J_0,\theta_0)$ lies in a periodic
orbit of period $p$ if $J_0p+\pi j(p-1)=2\pi s$ with $s$ integer,
and $\theta_0$ is arbitrary, as one can see from (\ref{I1}) and
(\ref{theta_n}). It is easy to find numerically values of $j,p,s$
such that none of the points in such an orbit is fixed under
$\invo_A$ or $\invo_B$: for instance, with $p=6$, $j=3$, and
$s=2$, it is easy to check
 that none of the fixed-point conditions (\ref{Jodd}),
(\ref{Jeven}) or (\ref{Jeven2}) is satisfied for any of the points
in  the periodic orbit.   Such an orbit can then be used to find
periodic orbits where none of the points is a fixed point of the
involutions, also for $\tilde{k}>0$. For this purpose take
$\tilde{k}>0$ but small. Iterate $p$ times the line of the fixed
points of the map with $\tilde{k}=0$, generating a new line
$M^p(J_0,\theta_0)$. Let $(\tilde{J}_0,\tilde{\theta}_0)$ be a
point of the intersection of these two lines. Although this is not
necessarily a fixed point of $M^p$ for $\tilde{k}>0$, it is
reasonable to look for one in its vicinity, and indeed we found
such orbits of period $p=6$. They are plotted  in Fig.
\ref{portp6inp2}. These orbits are also related by the symmetry
$J\longrightarrow -J$ discussed in Appendix \ref{appsymmetry}.

\begin{figure}
\begin{center}
\epsfxsize=13cm \leavevmode
    \epsffile{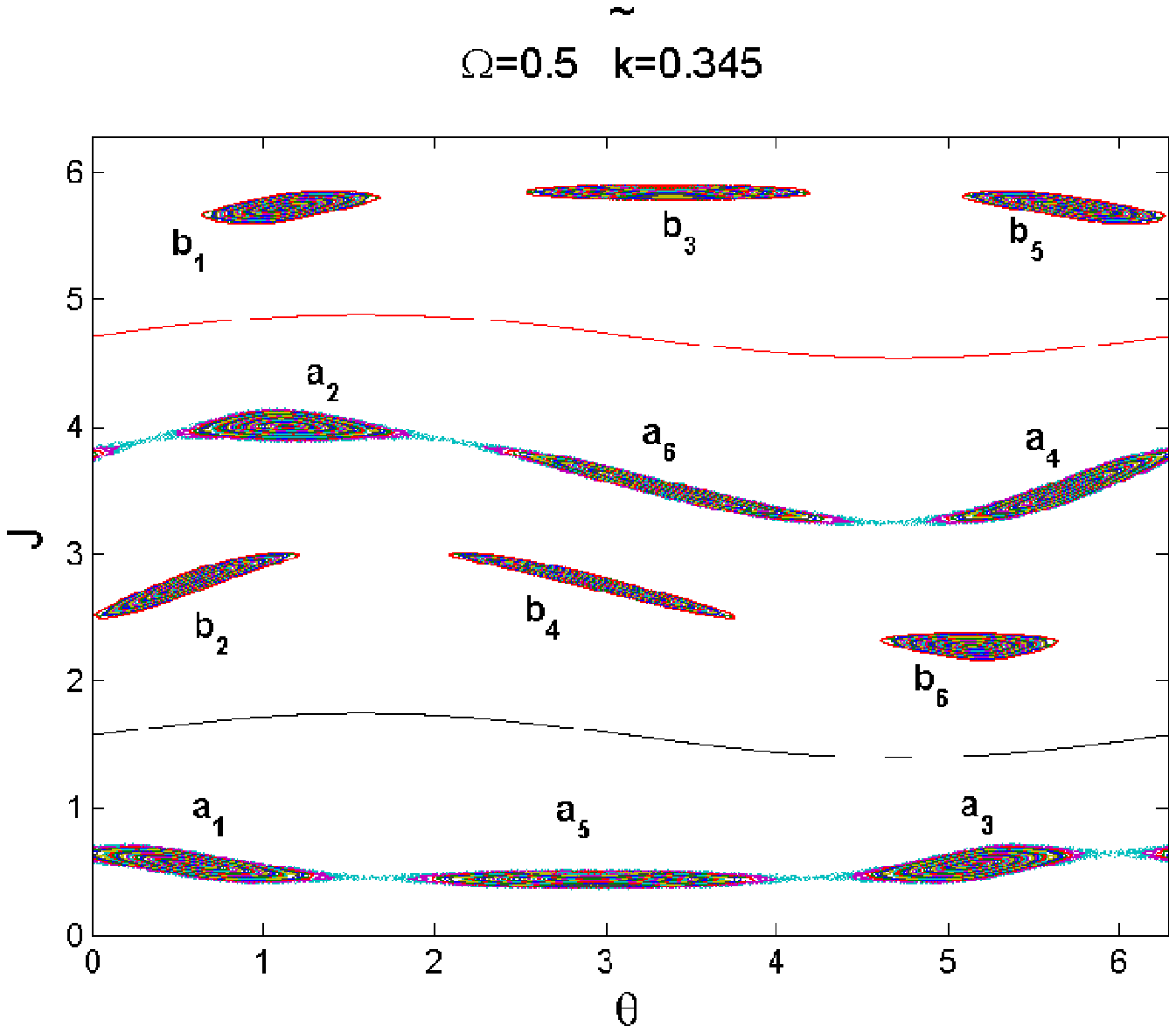}
\caption{The two period-$6$ orbits found at $\Omega=1/2$ for
$\tilde{k}=0.345$. The two lines indicate the invariant lines
under the $\invo_B$ involution (as given by (\ref{Jeven}) and
(\ref{Jeven2})), thus exhibiting that none of the points of these
periodic orbits is a fixed point of the $\invo_B$ involution. It
is also evident that none of the points is a fixed point of
$\invo_A$, since none of the points is located on the invariant
line of $\invo_A$ which is given by (\ref{Jodd}). Points in one
periodic orbit are marked by $a_i$ $i=1,2,...,6$ and points in the
other by $b_i$. The involution $\invo_A$ transforms the former
orbit into the latter, as follows:  $a_1\longrightarrow b_1$,
$a_2\longrightarrow b_6$, $a_3\longrightarrow b_5$,
$a_4\longrightarrow b_4$, $a_5\longrightarrow b_3$,
$a_6\longrightarrow b_2$, while $\invo_B$ transforms them as
follows: $a_1\longrightarrow b_2$, $a_2\longrightarrow b_1$,
$a_3\longrightarrow b_6$, $a_4\longrightarrow b_5$,
$a_5\longrightarrow b_4$, $a_6\longrightarrow b_3$. This is a
manifestation of theorem \ref{th3} of Appendix \ref{AppInv}.
}\label{portp6inp2}
\end{center}
\end{figure}

\begin{table}
 \centerline{
\begin{tabular}{||c|c|c|c|c|c|c||} \hline
tongue $(p,j)$ & $\Omega$ & $\tilde{k}_s$ & $\tilde{k}_{\infty}$ & $\alpha$ & $\delta$ & $n_{max}$\\
\hline $(1,0)$ & 0.3692 & $4.6239$ & $5.1290$ & $-4.0645$ &
$8.5839$ & $4$
\\
\hline $(2,1)$& $0.3624$ & $2.0019$ & $2.2790$ & $-3.9919$ &
$8.8307$ & $4$ \\ \hline $(2,1)$ & $0.6376$ & $2.0019$ & $2.2790$
&
$-3.9919$ & $8.8307$ & $4$ \\
\hline $(2,1)$ & $0.4878$ & $3.9126$ & $3.9711$ & $-4.0339$ &
$8.7034$ & $4$
\\ \hline $(2,1)$&
 $0.5122$ & $3.9126$ & $3.9711$ & $-4.0339$ & $8.7034$ & $4$ \\
\hline $(3,1)$ & $0.3936$ & $1.03$ & $1.1601 $ & $-4.1375$ &
$8.8534$ & $4$\\  \hline $(3,1)$ & $0.2258$ & $1.5169$ & $1.6167$
& $-4.0315$ & $9.0737$ & $4$ \\ \hline
\end{tabular}}
\caption{\label{tav2}As mentioned in the text, the stability of
orbits is lost through a period doubling cascade. In this table
some examples are given ,for different values of $\Omega$.}
\end{table}

\section{Summary and Discussion\label{summary}}

In this paper the tongues with $j/p=1, 1/2, 1/3$ were analyzed in
both the perturbative and the nonperturbative regime. A large
variety of scenarios for the disappearance of orbits, the loss of
their stability and bifurcations was found. We expect similar
behavior for the higher order tongues, but the verification is
left for further research. For higher values of $\tilde{k}$ some
novel results may be found but such an exploration is out of the
scope of the present paper.

The research in the present work was confined to the classical
regime. The fingerprints of the rich classical behavior that was
found on the quantum dynamics in the semiclassical regime may be
of great interest. In particular the effect of bifurcations on the
eigenstates or the scattering states of the quantum system may
reveal novel results. Effects similar to the ones found for
quantum ratchets \cite{Ratchets} and "flooding" \cite{Ketzmerick}
are expected. Also states on islands around periodic orbits
related by symmetry may result in states of novel properties.
These may lead to effects observable in atom optics experiments.

\begin{acknowledgments} This work was partly supported by the Israel
Science Foundation (ISF), by the US-Israel Binational Science
Foundation (BSF), by the Minerva Center of Nonlinear Physics of
Complex Systems, by the Shlomo Kaplansky academic chair and by the
Institute of Theoretical Physics at the Technion. Highly
instructive discussions with Roberto Artuso, Andreas Buchleitner,
Michael d'Arcy, Simon Gardiner, Shai Haran, Zhao-Yuan Ma, Zeev
Rudnick, Gil Summy, are acknowledged. I.G. acknowledges partial
support from the MIUR-PRIN project "Order and Chaos in extended
nonlinear systems: coherent structures, weak stochasticity and
anomalous transport".
\end{acknowledgments}

\begin{appendix}
\section{The Involution Method\label{AppInv}}
\def\theequation{A. \arabic{equation}}
\setcounter{equation}{0}

This method allows for establishing  an analytical  relation
between the coordinates $J,\theta$ of a point on a periodic orbit,
and in this way it considerably simplifies the task of computing
periodic orbits. The method was introduced by Greene \cite{Greene}
who implemented it in the computation of the approximants of the
last bounding KAM trajectory of the Standard Map. In this Appendix
the method is reviewed for the reader's convenience, and its
application to the map $\map$ of (\ref{I1}) is formulated.

Consider a map that can be written as a product of two
involutions, i.e.
\begin{equation}\label{Mproduct}
\map\;=\;\invo_2\;\invo_1\;,
\end{equation}
where the maps $\invo_1$ and $\invo_2$ satisfy
$\invo_1^2=\invo_2^2=1$ (the identity map). The following theorems
hold:

\begin{theo}
\label{th1} {If, for some point $X_0$ in phase space, and some
integer $N$,
\begin{equation}\label{TheoremI1}
\invo_1\;(X_0)\;=\;X_0\;,
\end{equation}
and
\begin{equation}\label{TheoremI2}
\invo_1\;(\map^N\;(X_0))\;=\;\map^N\;(X_0)\;,
\end{equation}
then
\begin{equation}\label{TheoremI3}
\map^{2N}\;(X_0)\;=\;X_0\;.
\end{equation}
} \end{theo}
\begin{theo}
\label{th2} { If, for some point $X_0$ in phase space, and some
integer $N$,
\begin{equation}\label{TheoremII1}
\invo_2\;(X_0)\;=\;X_0\;,
\end{equation}
and
\begin{equation}\label{TheoremII2}
\invo_1\;(\map^N\;(X_0))\;=\;\map^N\;(X_0)\;,
\end{equation}
then
\begin{equation}\label{TheoremII3}
\map^{2N+1}\;(X_0)\;=\;X_0\;.
\end{equation}
} \end{theo} \noindent
 These theorems are easily proven by
explicit use of (\ref{TheoremI1}) and (\ref{TheoremI2}) in the RHS
of (\ref{TheoremI3}), and of (\ref{TheoremII1}), and of
(\ref{TheoremII2}) in the RHS of (\ref{TheoremII3}).

The opposite direction is not correct. Orbits can be found, such
that no point in them is fixed under $\invo_1$ or $\invo_2$ (see
section \ref{noinvo}).


\begin{theo}\label{th3}
{Let $\orb\equiv\{X_1, X_2,.... X_p\}$ be a primitive periodic
orbit of period $p$ of the map $\map$. For $i=1,2$ denote
$\orb^{(i)}\equiv \invo_{i} (\orb)$. Then $\orb^{(2)}$ is a
primitive periodic orbit of period $p$, and
$\orb^{(1)}=\orb^{(2)}$.}
\end{theo}

{\em Proof}: From (\ref{Mproduct}) it follows that $\map\invo_2=\
\invo_2\map^{-1}$. Therefore, for any $X_j\in\orb$,
\begin{equation}
\label{orbit12}
\map(\invo_2(X_j))\;=\;\invo_2(\map^{-1}(X_j))\in\;\invo_2(\orb)\;=\;\orb^{(2)}\;,
\end{equation}
hence $\map(\orb^{(2)})\subseteq\orb^{(2)}$, and then
$\map(\orb^{(2)})=\orb^{(2)}$ follows, because $\map$ is
one-to-one. Furthermore, (\ref{orbit12}) entails that $\orb^{(2)}$
is the orbit of any of the points in it, because this is true of
$\orb$. Finally,
\begin{eqnarray}
\orb^{(1)}\;&=&\;\invo_1(\orb)\;=\;\invo_1\invo_2\invo_2(\orb)\nonumber\\
&=&\;\invo_1\invo_2(\orb^{(2)})\;=\; \map^{-1}(\orb^{(2)})\nonumber\\
&=&\;\orb^{(2)}\;.
\end{eqnarray}
Theorem \ref{th3} has the following direct
\par\noindent
{\bf Corollary:} If in $\orb$ there is a fixed point of either
involution, then $\orb=\orb^{(1)}=\orb^{(2)}$, and so the orbit is
invariant under both involutions. The orbit can be divided in
classes such that each class either contains one fixed point of
one involution or it contains two points which are mapped onto
each other by that involution. In the case when the involution is
$\invo_A$, points in each such  pair have opposite $J$ . For odd
$p$, this implies that a point with $J=0$ belongs to the periodic
orbit (and is a fixed point of $\invo_A$).


Theorem \ref{th1} is employed for finding periodic orbits of even
period and Theorem \ref{th2} for finding periodic orbits of odd
period \cite{Greene}. One solves for periodic orbits, assuming
that in a periodic orbit there is a fixed point $X_0$ of either
$\invo_A$ or $\invo_B$ (in our case, $X_0=(J_0,\theta_0)$). This
assumption reduces the number of unknowns. Since the periodic
orbits of $\map$
 and of $\map^{-1}$ are identical, and $\map^{-1}=\invo_1\invo_2$
 holds, we can
interchange $\invo_A$ and $\invo_B$ in the calculation of periodic
orbits. For orbits of odd period one can easily check that
$X_0=(J_0=0,\theta_0)$ (eq.~\ref{Jodd}) is a fixed point of
$\invo_A$. In our applications periodic orbits with an odd period
$p$ contain a fixed point of $\invo_A$ where (\ref{Jodd}) is
satisfied, while periodic orbits with an even period typically
contain a fixed point of $\invo_B$, where (\ref{Jeven}) or
(\ref{Jeven2}) is satisfied. By the corollary, these orbits must
contain more than just 1 fixed point of the latter involution.

\section{Time-reversal symmetry.}\label{appsymmetry}
\def\theequation{B. \arabic{equation}}
\setcounter{equation}{0}  In this appendix a symmetry of periodic
orbits of $\map$ is discussed, which is related to time-reversal.
The map $\map$ may be written as $\map=\map_K\map_F$, where
\begin{eqnarray}\label{symm3}
\map_F\;&:&\;(J,\theta)\;\rightarrow\;(J,\;\theta+J)\;,\nonumber\\
\map_K\;&:&\;(J,\theta)\;\rightarrow\;(J+{\tilde
k}\sin(\theta)+2\pi\Omega,\theta)\;.
\end{eqnarray}
(throughout this appendix, mod $2\pi$ is left understood in maps).
 Let $\map'\equiv\map_F\map_K$. The map $\map'$ is easily seen to
satisfy:
$$
\map^{-1}\;=\;\invo\,\map'\,\invo\;
$$
where
\begin{equation} \label{I}
\invo\;:\;(J,\theta)\;\rightarrow\;(-J,\theta)\;.
\end{equation}
and so, denoting ${\mathfrak R}\equiv \invo\map_F$, the map
$\mathfrak R$ is an involution: $\rmap^2=1$, and satisfies
\begin{equation}
\label{trev} \map^{-1}\;=\;{\mathfrak R}\,\map\,{\mathfrak R}\;.
\end{equation}
Therefore $\mathfrak R$ achieves the time-reversal of the map
$\map$. We have thus proven that:
\begin{propo}\label{trevsym}
 {$\map$ is time-reversal invariant, in the sense that an
area-preserving involution $\mathfrak R$ exists, such that ~
${\mathfrak R}\map{\mathfrak R}=\map^{-1}$.~ Therefore, if $\orb$
is a periodic orbit of $\map$, then the time-reversed orbit
${\mathfrak R}(\orb)$ is also a periodic orbit of $\map$, and the
two orbits have the same stability.}
\end{propo}
{\bf Corollary:} If for the map (\ref{I1}) there is only one
periodic orbit of a given stability (as determined by the tangent
map), then $\orb={\mathfrak R}(\orb)$ must hold,  and so for each
point in the orbit with momentum $J$ there is another point  with
the value $-J$.

\section{{Gauss Sums and periodic orbits}\label{appgauss}
}
\def\theequation{C. \arabic{equation}}
\setcounter{equation}{0} In this appendix some properties of Gauss
sums that are relevant for the present paper will be reviewed.
Specifically we focus on the sum (\ref{Gpj}), that is related to
\begin{equation}\label{Gjpl}
\tilde{G}(j,p,l)=\frac{1}{p}\sum_{m=1}^{p}e^{\frac{i\pi}{p}[lm+jm^2]}
\end{equation}
where in our case $l=j(\chi_p-1)$ vanishes for even p if
(\ref{Jeven}) holds and it is equal to $p$ if (\ref{Jeven2}) holds
while it takes the value $-j$ for odd $p$. Using elementary
calculations one can show \cite{GRF} that in our case (\ref{Gjpl})
is just (\ref{Gpjcompact}) divided by $p$, namely
\begin{equation}\label{GpG}
G(j,p)=p\tilde{G}(j,p,l).
\end{equation}
The calculation of $\xi(j,p)$ is more complicated and will be
summarized in what follows. In the standard number theoretical
literature (\cite{Gauss sums2} \cite{Dirichlet}), sums like
(\ref{Gjpl}) are calculated for $l=0$, even $j$ and odd $p$.
Hannay and Berry \cite{Gauss sums} generalized to other cases of
(\ref{Gjpl}) (See also \cite{talbot}). They show that this sum
reduces to the average of the summands and is given, for $j$ and
$p$ mutually prime, by Eq. (14) there, which reads:

\begin{eqnarray}\label{14}
\tilde{G}(j,p,l) = \left\{ \begin{array}{ll} \frac{1}{\sqrt{p}}
\left( \begin{array}{c} j \\ p
\end{array} \right) \exp \left[ -\frac{i\pi}{4}(p-1)-\frac{i\pi j}{p}(j\backslash p)^2(\frac{l}{p})^2 \right] & \mbox{$p$ odd, $j$ even, $l$ even}
\\ \\
\frac{1}{\sqrt{p}} \left( \begin{array}{c} j \\ p
\end{array} \right)
\exp \left[ -\frac{i\pi}{4}(p-1)-\frac{i4\pi j}{p}(4j\backslash
p)^2l^2 \right] & \mbox{$p$ odd, $j$ odd, $l$ odd}
\\ \\
\frac{1}{\sqrt{p}} \left( \begin{array}{c} p \\ j
\end{array} \right) \exp \left[ \frac{i\pi}{4}j-\frac{i\pi j}{p}(j\backslash p)^2(\frac{l}{p})^2 \right] & \mbox{$p$ even, $j$ odd, $l$ even}

\end{array} \right.
\end{eqnarray}
where $(j\backslash p)$ is the inverse (mod $p$) of $j$, notably
\begin{equation}\label{inversej}
j(j\backslash p)\;=\;1 ~\mbox{mod}\;p
\end{equation}
which can be calculated with the help of (whenever $j$ is mutually
prime with $p$):
\begin{equation}
(j\backslash p)=j^{\varphi(p)-1} ~\mbox{mod $p$} \end{equation}
where $\varphi(p)$ is Euler's totient function, that is, the
number of integers less than $p$ that have no common devisors with
$p$.
$ \left( \begin{array}{c} a \\
b
\end{array} \right)$ is the Jacobi symbol, where $a$ and $b$ are integers so that $b$ is {\em
odd} and does not divide $a$. The Jacobi symbol assumes the values
$\pm1$ only, and is the
product of the Legendre symbols $ \left( \begin{array}{c} a \\
q
\end{array} \right)$ for all the {\em prime} factors $q$ of $b$. The Legendre symbol is defined by
\begin{equation}
\left( \begin{array}{c} a \\
q
\end{array} \right) \equiv \left\{ \begin{array}{ll}
~1 & \mbox{if there is an integer $m$ such that $m^2=a~$ mod $q$} \\
-1 & \mbox{if there is no integer $m$ such that $m^2=a~$ mod $q$}
\end{array} \right.
\end{equation}
A more straightforward way of determining the value of a Legendre
symbol is given by the next formula, which is based on Fermat's
little theorem :
\begin{equation}
\left( \begin{array}{c} j \\ p \end{array} \right) =
j^{\frac{p-1}{2}} ~\mbox{mod $p$}
\end{equation}
keeping in mind that $p$ is an {\em odd prime} and that $j$ is not
divisible by $p$. For our purposes it is convenient to write the
Jacobi symbol in the form
\begin{equation}
 \left( \begin{array}{c}
j \\ p
\end{array}  \right) = \exp \left[ \frac{i\pi}{2} \left( 1- \left( \begin{array}{c}
j \\ p
\end{array}  \right) \right) \right].
\end{equation}
It is useful to note that
\begin{equation}
\left( \begin{array}{c} -j \\ p \end{array} \right) = \left(
\begin{array}{c} j \\ p \end{array} \right) \left( \begin{array}{c} -1 \\ p \end{array} \right)
\end{equation}
and that
\begin{equation}
\left( \begin{array}{c} -1 \\ p \end{array} \right) =
e^{\frac{i\pi}{2}(p-1)}
\end{equation}
and also that
\begin{equation}
\left( \begin{array}{c} j \\ 1  \end{array} \right) = 1
\end{equation}
for all integers $j$. Other useful formulae are given in appendix
B of \cite{Gauss sums} and in \cite{Gauss sums2} and
\cite{Dirichlet} (see also \cite{talbot}).

We note that in our case, for even $p$, $l=0$ in case
(\ref{Jeven}), and $l=p$ in case (\ref{Jeven2}), while for odd
$p,~~ l=-j$. In this case (\ref{Gjpl}) takes the form
\begin{equation}\label{Gjpl2}
\tilde{G}(j,p,l)=\frac{1}{\sqrt{p}}~e^{i\xi(p,j)}
\end{equation}
where, for even $p$ corresponding to (\ref{Jeven}),
\begin{equation}\label{peven}
\xi(p,j)= \frac{\pi}{2} \left[ 1- \left( \begin{array}{c} p \\ j
\end{array} \right) + \frac{j}{2} \right],
\end{equation}
and for even $p$ corresponding to (\ref{Jeven2})
\begin{equation}\label{peven2}
\xi(p,j)= \frac{\pi}{2} \left[ 1- \left( \begin{array}{c} p \\ j
\end{array} \right) - 2\frac{j}{p}(j\backslash p)^2 \right],
\end{equation}
while for odd p
\begin{equation}\label{podd}
\xi(p,j)= \left\{ \begin{array}{ll} \frac{\pi}{2} \left[ 1- \left(
\begin{array}{c} j \\ p
\end{array} \right)
 -\frac{1}{2}(p-1)-
\frac{1}{2}\frac{j}{p}(j\backslash p)^2l^2 \right] & \mbox{$j$ even} \\ \\
\frac{\pi}{2} \left[ 1- \left( \begin{array}{c} j \\ p
\end{array} \right) -\frac{1}{2}(p-1)-
\frac{8j}{p}(4j\backslash p)^2l^2 \right] & \mbox{$j$ odd}
\end{array} \right .
\end{equation}

The phase $\xi(p,j)$ determines the relation between $\theta_0$
and $\vartheta$ via (\ref{theta3def}) and can be used to check the
accuracy of the calculations. At the center of the tongue
$\vartheta=0$ or $\vartheta=\pi$ while at the boundaries
$\vartheta=\frac{\pi}{2}$ or $\vartheta=\frac{3\pi}{2}$. In table
\ref{ksitable} we compare the values of the numerical results
$\Delta\theta=\vartheta -\theta_0$ 
for $\tilde{k}=0.001$ and $\theta_o$ at the center of some tongues
with $\xi(p,j)$ calculated from (\ref{peven}) and (\ref{podd}).


\begin{table}\label{ksitable}
\centerline{
\begin{tabular}{||l|c|c||} \hline
tongue $(p,j)$& $\Delta\theta=\vartheta -\theta_0$ & $\xi(p,j)$ \\
\hline $(2,1)$ & 0.7855 & $\pi /4\approx 0.7853$ \\ \hline $(3,1)$
& 0.5235 & $\pi/6\approx 0.5235$ \\ \hline $(3,2)$ & -0.5235 &
$-\pi/6\approx -0.5235$
\\ \hline
$(4,1)$ & 0.7855 & $\pi/4\approx 0.7853$
\\ \hline
$(4,3)$ & 2.3567 & $3\pi/4\approx 2.3561$
\\ \hline
$(5,1)$ & 0.6284 & $\pi/5\approx 0.6283$
\\ \hline
$(5,2)$ & 1.2573 & $2\pi/5\approx 1.2566$
\\ \hline
$(5,3)$ & -1.2573 & $-2\pi/5\approx -1.2566$
\\ \hline
$(5,4)$ & -0.6284 & $-\pi/5\approx -0.6283$
\\ \hline
$(6,1)$ & 0.7857 & $\pi/4\approx 0.7853$
\\ \hline
$(6,5)$ & -2.3558 & $-3\pi/4\approx -2.3561$
\\ \hline
$(7,1)$ & 0.6733 & $3\pi/14\approx 0.6731$
\\ \hline
$(7,2)$ & -0.2251 & $-\pi/14\approx -0.2243$
\\ \hline
$(7,3)$ & 2.0209 & $9\pi/14\approx 2.0195$
\\ \hline
$(7,4)$ & -2.0209 & $-9\pi/14\approx -2.0195$
\\ \hline
$(7,5)$ & 0.2251 & $\pi/14\approx 0.2243$
\\ \hline
$(7,6)$ & -0.6733 & $-3\pi/14\approx -0.6731$
\\ \hline
\end{tabular}} \caption{\label{ksitable}Numerically computed
values of $\Delta\theta=\vartheta-\theta_0$, compared to
$\xi(p,j)$ (see (\ref{theta3def})).}
\end{table}


\section{A numerical method for the calculation of periodic orbits \label{appmatlab}}
\def\theequation{D. \arabic{equation}}
\setcounter{equation}{0} In the numerical solution for periodic
orbits we make use of (\ref{Jodd}), (\ref{Jeven}) and
(\ref{Jeven2})  in order to eliminate $J_0$. Then we impose the
periodic orbit conditions (\ref{theta}) and require that
$\theta_0$ is a
common zero of the functions 
\begin{equation}\label{C1}
F_J(\theta_0)=\tilde{k}\sum_{n=1}^p\sin\theta_n+2\pi\Omega p -2\pi
j,
\end{equation}
where $j$ is determined by the tongue ((\ref{theta}) and
(\ref{Omega_0})), and
\begin{equation}\label{C2}
F_{\theta}(\theta_0)=pJ_0+\tilde{k}\sum_{n=1}^{p}(p-n)\sin\theta_{n}+
\pi\Omega p(p-1)-2\pi s \, ,
\end{equation}
that are obtained after iterating the map (\ref{I1}) $p$ times. We
solved for zeros of $F_J(\theta_0)$ and $\sin
[\frac{1}{2}F_{\theta}(\theta_0)]$. This is because $j$ is known,
while $s$ is unknown and since the function $F_{\theta}(\theta_0)$
should be taken mod $2\pi$. Then only the common zeros of
$F_J(\theta_0)$ and $\sin [\frac{1}{2} F_{\theta}(\theta_0)]$ are
kept. The elimination of $J_0$ by the involution method, described
in Appendix \ref{AppInv}, reduced the problem to one of finding
zeroes of a function of a single variable, that is a relatively
easy task. We used the MATLAB7 routine ``fzero'' for this purpose
\cite{MATLAB}. In order to locate the boundary, we note that the
zeroes corresponding to the stable and unstable orbits are
separated by an extremum of the function in question, say
$F_J(\theta_0)$. At the boundary these zeroes coalesce and at this
extremum the relevant function vanishes (at the extremum $F_J=0$).
The extremum can be easily found by the MATLAB7 routine
``fminbnd'' \cite{MATLAB}. The boundary of a tongue in the
$\tilde{k},\Omega$ space is the locus of points of vanishing
extremum. It was found by successive change of $\tilde{k}$ and
$\Omega$ starting from vertex $\tilde{k}=0,~ \Omega=j/p$. The
boundary was found by the overshoot method, making use of the fact
that the value of the function at the extremum changes sign when
the boundary of a tongue is crossed.


\section{The resonance of order 3 for the $p=1$
tongue} \label {apphopfbifu}
\def\theequation{E. \arabic{equation}}
\setcounter{equation}{0}

When the parameter $\tilde{k}$ is increased across the stability
border, the typical scenario of a Hopf bifurcation is observed,
with period doubling accompanied by the birth of a resonance of
order 3, as shown in Fig. \ref{port1}. Here we obtain analytical
expressions for the first period doubling, and the normal form for
the resonance of order 3.

For the map (\ref{I1}) that is studied in the present work, the
fixed point $(J_0,\theta_0)$, discussed in subsection \ref{IIIA},
loses stability for $\tilde{k}_0\cos\theta_0=-4$. In order to
analyze the resulting bifurcation, we study the small deviations
\begin{eqnarray}\label{nf1}
\delta\theta&=&\theta-\theta_0 \nonumber \\
\delta J&=& J-J_0\, .
\end{eqnarray}
From (\ref{I1}) we find that to the second order in
$\delta\theta$,
\begin{eqnarray}\label{nf2}
\delta{J}_{n+1}&=&\delta J_n+A\delta\theta_n+D(\delta\theta_n)^2
\nonumber \\
\delta\theta_{n+1}&=&\delta\theta_n+\delta J_{n+1}\, ,
\end{eqnarray}
where $A=\tilde{k}\cos\theta_0,~D=-(\tilde{k}/2)\sin\theta_0$. It
is convenient to introduce the variable $x=(D/2)\delta\theta$ and
turn Eqs. (\ref{nf2}) into one second order standard equation (see
\cite{Lichtenberg_Lieberman}, p. 243)
\begin{equation}\label{nf3}
x_{n+1}+x_{n-1}=2Cx_n+2x_n^2\, ,
\end{equation}
where $C=(2+A)/2$. In our case we are interested in the vicinity
of $A=-4$, or  $C=-1$. For $C>-1$, the fixed point $x_n=0$
(corresponding to the original point $(J_0,\theta_0)$) is stable.
For $C<-1$, it becomes unstable and a stable periodic orbit which
consists of the points
\begin{equation}\label{nf4}
x_{2\pm}=a\pm b\, ,
\end{equation}
where $a=-(C+1)/2$ and $b=\frac{1}{2}\sqrt{(C+1)(C-3)}$ appears.
If $C=-1-\eta$, then $a=\eta/2$ and
$b=\frac{1}{2}\sqrt{\eta(2+\eta)}$. To study the small deviations,
we denote
\begin{equation}\label{nf5}
x_n=x_{2\pm}+\Delta x_n\, ,
\end{equation}
where we numerate the steps so that for even $n$ the point $x_n$
is near $x_{2+}$, while for odd $n$ it is near $x_{2-}$. Iterating
the map and keeping terms to the second order in the $\Delta x_n$,
one finds for even $n$
\begin{equation}\label{nf6}
q_{n+2}+q_{n-2}=2C^{\prime}q_n+2q_n^2\, ,
\end{equation}
where
\begin{equation}\label{nf7}
q_n=2[(C+2x_{2-})+(C+x_{2+})^2]\Delta x_n
\end{equation}
and $C^{\prime}=-2C^2+4C+7$. In our case
$C^{\prime}=1-8\eta-2\eta^2$. For odd $n$ the same result is
found, but with $x_{2-}$ and $x_{2+}$ interchanged.

The map (\ref{nf6}) can be obtained from the Hamiltonian
\begin{equation}\label{nf8}
H=\frac{p^2}{2}+V(q)\sum_{n=-\infty}^{\infty}\delta(t-n)
\end{equation}
with the time rescaled by the factor of $2$ and
\begin{equation}\label{nf9}
V(q)=-\frac{1}{2}A^{\prime}q^2+\frac{2}{3}q^3
\end{equation}
with $A^{\prime}$ related to $C^{\prime}$ via
$C^{\prime}=(A^{\prime}+2)/2$. To see this, one notes that the map
obtained from the Hamiltonian (\ref{nf8}) is similar to Eq.
(\ref{nf2}), and therefore, Eq. (\ref{nf6}) can be obtained from
Eq. (\ref{nf8}) in the same way  as Eq. (\ref{nf3}) is obtained
from Eq. (\ref{nf2}). The parameter  that controls the potential
is
\begin{equation}\label{nf10}
A^{\prime}=2C^{\prime}-2=-4C^2+8C+5=-16\eta-4\eta^2\, .
\end{equation}
Using the Poisson summation formula we write the periodic $\delta$
function in the form $\sum_{n=-\infty}^{\infty}\delta(t-n)=1+
2\sum_{n=1}^{\infty}\cos(2\pi n t)$. For small $q$ the averaged,
or time independent, part reduces to a harmonic oscillator with
the frequency $\omega=\sqrt{-A^{\prime}}=2\sqrt{4\eta+\eta^2}$.
The third order resonance is found for $\omega=2\pi/3$.
Introducing the action--angle variables $(I,\phi)$ for the linear
oscillator part, where $q=\sqrt{2I/\omega}\cos\phi$ and
$p=\sqrt{2I\omega}\sin\phi$, the Hamiltonian (\ref{nf8}) reads
\begin{equation}\label{nf11}
H=\omega
I+\frac{4}{3}\left(\frac{I}{2\omega}\right)^{3/2}\cos(3\phi-2\pi
t)+ {\cal R}(I,\phi,t)\, ,
\end{equation}
where ${\cal R}(I,\phi,t)$ is the non-resonant part of the
Hamiltonian (\ref{nf8}). Introducing the resonant phase
$\psi=\phi-2\pi t/3$ and the detuning from the linear resonance
$\delta=\omega-2\pi/3$, and denoting $r=(4/3)/(2\omega)^{3/2}$ we
obtain the normal form for the 3--resonance
\cite{arnold1,arnold2}:
\begin{equation}\label{nf14}
H_3=\delta\cdot I+rI^{3/2}\cos(3\psi)\, .
\end{equation}
Denoting $\bar{x}=\sqrt{I}\cos(\psi),~\bar{y}=\sqrt{I}\sin(\psi)$,
we obtain another well known expression for the normal form of the
resonance of the order of 3 \cite{arnold1,arnold2}
\begin{equation}
H_3=\delta(\bar{x}^2+\bar{y}^2)+r(\bar{x}^3-3\bar{x}\bar{y}^2)\, .
\end{equation}
As $\tilde{k}$ increases the orbit (\ref{nf4}) loses its stability
and bifurcates, typically via a period doubling bifurcation
cascade.

\end{appendix}


\end{document}